\documentclass[12pt]{iopart}
\usepackage{graphicx}
\usepackage{iopams}  
\usepackage[square, comma, sort&compress, sectionbib,numbers]{natbib}
\usepackage{verbatim}

\def\newblock{\hskip .11em plus .33em minus .07em}

\begin{document}
\title[Imaging universal conductance fluctuations in graphene]{Imaging coherent transport in graphene (Part I): Mapping universal conductance fluctuations}
\author{J Berezovsky$^1$, M F Borunda$^2$, E J Heller$^2$ \& R M Westervelt$^1$}
\address{$^1$ School of Engineering and Applied Science, and Department of Physics, \\Harvard University, Cambridge, MA 02138.\\ $^2$ Department of Chemistry and Chemical Biology, and Department of Physics, Harvard University, Cambridge, MA 02138.}
\ead{westervelt@seas.harvard.edu}

\begin{abstract}
Graphene provides a fascinating testbed for new physics and exciting opportunities for future applications based on quantum phenomena.  To understand the coherent flow of electrons through a graphene device, we employ a nanoscale probe that can access the relevant length scales - the tip of a liquid-He-cooled scanning probe microscope (SPM) capacitively couples to the graphene device below, creating a movable scatterer for electron waves.  At sufficiently low temperatures and small size scales, the diffusive transport of electrons through graphene becomes coherent, leading to universal conductance fluctuations (UCF).  By scanning the tip over a device, we map these conductance fluctuations \textit{vs.} scatterer position.  We find that the conductance is highly sensitive to the tip position, producing $\delta G \sim e^2/h$ fluctuations when the tip is displaced by a distance comparable to half the Fermi wavelength.  These measurements are in good agreement with detailed quantum simulations of the imaging experiment, and demonstrate the value of a cooled SPM for probing coherent transport in graphene.
\end{abstract}

\maketitle

\section{Introduction}
Graphene, a single atomic layer of carbon in a hexagonal lattice, has remarkable properties.  It has conical conduction and valence bands that meet at a single point in $k$-space (the Dirac point)~\cite{Geim:2007}.  Strong quantum confinement effects have been observed in quantum dots and nanoribbons~\cite{Han:2007}, and the quantum Hall effect can be seen at room temperature~\cite{Novoselov:2007}.  Scanning tunneling microscopy has measured the surface topography~\cite{Stolyarova:2007}, local charge density~\cite{Zhang:2009}, and the local density of states~\cite{Rutter:2007, Deshpande:2009}, and a scanned charge sensor has been used to map the charge density~\cite{Martin:2008}. 

Universal conductance fluctuations (UCF) ~\cite{Lee:1985, Altshuler:1985,Washburn:1986} occur when a coherent electron wave scatters repeatedly while it travels through a disordered conductor, following all possible paths through the sample.  The different paths interfere with each other, creating a change in the conductance known as UCF that depends sensitively on the scatterer positions.  When the size of the sample is less than the diffusive phase coherence length $L_\phi$, interference between paths yields a universal magnitude $\delta G \sim e^2/h$ for UCF, independent of the sample size and the degree of disorder.  Theory~\cite{Feng:1986, Altshuler:1985b} has predicted that the full UCF effect is obtained by moving a single scatterer a distance comparable to the Fermi wavelength $\lambda_F$. 

In this work, we use a liquid-He-cooled scanning probe microscope (SPM)~\cite{Tessmer:1998,Topinka:2000,Topinka:2003,Pioda:2004,Jura:2007, Aidala:2007,Martin:2008,Braun:2008}  to study coherent transport in graphene. We obtain conductance images that map the effect of a single scatterer on UCF.  A charged SPM tip near the surface of a graphene sample creates an image charge that acts as a movable scatterer. This alters the electron wave function in the vicinity of the tip, leading to changes in quantum interference that give rise to UCF.  An image of the sample conductance \textit{vs.} tip position provides a spatial ``fingerprint'' that is unique to the arrangement of scatterers at a given Fermi energy.  This technique allows us to observe the signatures of UCF without varying any external parameters (\textit{e.g.} the magnetic field or gate voltage).  Our approach reveals how UCF are created by the displacement of a single scatterer, as predicted by theory~\cite{Feng:1986, Altshuler:1985b}.  

UCF has recently been investigated in transport measurements of mesoscopic graphene samples~\cite{Graf:2007,Rycerz:2007,Heersche:2007,Staley:2008,Kechedzhi:2009,Horsell:2009}. Our SPM technique provides a valuable spatial probe of coherent transport in graphene:  1) The tip can be adjacent to the two-dimensional electron gas, maximizing the spatial resolution, because graphene is two-dimensional material; 2) The Fermi energy $E_F$ can be continuously varied from positive values for electrons, through zero to negative values for holes by using a back gate; 3) At $T=4$~K, the observed coherence length ($L_\phi \sim 500$~nm) and elastic mean free path ($l_e \sim 50$~nm) allow measurement in the coherent regime. Our results for graphene should also apply to other two-dimensional conductors, though some questions may be raised by scattering in graphene's unusual band structure~\cite{Rycerz:2007,Kharitonov:2008}.    

\section{Experimental Methods}
 \begin{figure}
 \centering
\includegraphics[width=0.7\textwidth]{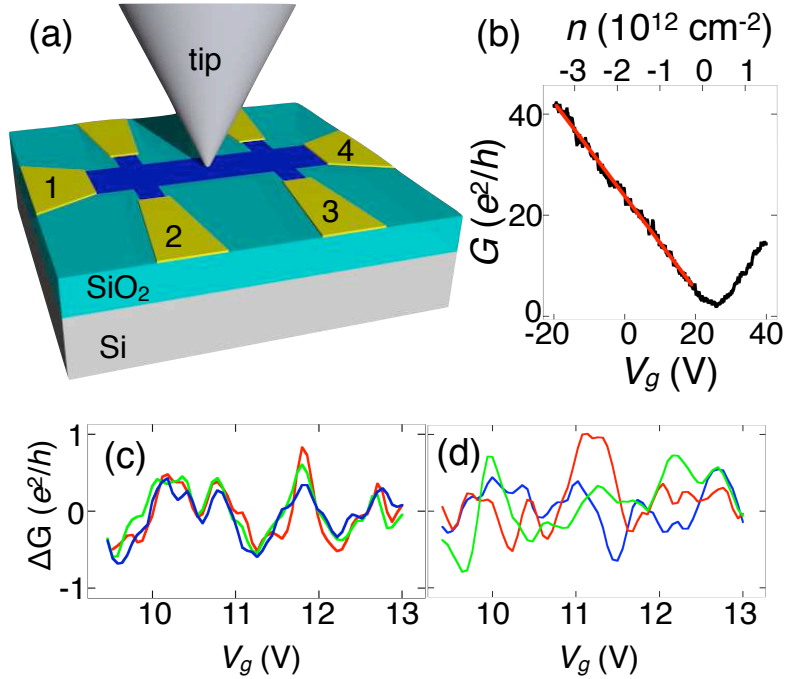}
\caption{(a) Schematic of the graphene sample mounted in the cooled scanning probe microscope (SPM), showing a Hall bar contacted by six Cr/Au leads.  The SPM tip and lead 4 are grounded, with a 25 nA rms current between leads 1 and 4 at 5 kHz.  Voltage is measured between leads 2 and 3 using a lock-in amplifier.  A back-gate voltage $V_g$ is applied to the degenerately doped Si substrate. (b) Measured conductance $G$ \textit{vs.} $V_g$, with a linear fit for $V_g <20$~V.  (c) Three consecutive measurements of $G$ \textit{vs.} $V_g$ with the tip fixed far from the sample, with the linear background from (b) subtracted.  (d) Same as (c), but with the tip 10~nm above the sample, at three different locations spaced $100$~nm apart. \label{fig:setup}}
 \end{figure}

The graphene samples studied in these experiments are single-atomic-layer Hall bars, with a geometry shown schematically in figure~\ref{fig:setup}a.  The experimental data in the figures below are from a sample with width 500~nm and voltage contacts (leads 2 and 3 in figure~\ref{fig:setup}a) with centers spaced 1200~nm apart.  Graphene flakes were prepared through mechanical exfoliation (the ``sticky tape method'') and deposited onto a degenerately doped Si substrate capped with 280~nm of SiO$_2$. A back gate voltage $V_g$ is applied between the substrate and the graphene to vary $E_F$ and the charge density $n$.  Using electron beam lithography, Cr/Au leads are deposited onto the graphene, after which the Hall bar structure is formed via an oxygen plasma etch.  The presence of single-layer graphene is confirmed by observing quantum Hall conductance plateaux at the expected values $4(\nu+1/2)e^2/h$, where $\nu$ is an integer.  The sample is then mounted on a home-built scanning gate microscope~\cite{Topinka:2000, Aidala:2007} and cooled in He exchange gas in thermal contact with a liquid He bath at $T=4.2$~K.  Similar results have been obtained on three other samples with similar dimensions.  The four-probe conductance $G$, shown in figure~\ref{fig:setup}b, displays the characteristic linear variation of $G$ \textit{vs.} $V_g$ on either side of the Dirac point at $V_{Dirac}=22$~V.  

Coherent, diffusive transport is expected when the sample size $L\lesssim L_\phi$, the electron's diffusive coherence length, and $L\gg l_e$, the elastic mean free path.  At $T= 4$~K, we obtain $L_\phi\approx 0.5~\mu$m from weak localization and magnetoconductance measurements (not shown).  $L_\phi$ is larger than the $0.4 \times 0.4~\mu$m$^2$ field of view in figure~\ref{fig:images}, and comparable to the sample width ($W=0.5~\mu$m) and length ($L=1.2~\mu$m); all of these lengths are much larger than $l_e \sim 50$~nm. 

From the slope of $G$ \textit{vs.} $V_g$ in figure~\ref{fig:setup}b and the capacitance between the back gate and the graphene, discussed below, the electron and hole mobility away from the Dirac point is found to be $\mu \approx 7200~\mathrm{cm}^2/\mathrm{Vs}$.  The shift of the Dirac point from $V_g=0$ to $V_{Dirac} = 22$~V is attributed to charged impurities located either above or below the graphene layer, which induce a charge in the graphene.

To create a movable scatterer, a conducting, voltage-biased SPM tip with radius of curvature $r_{tip}=20$~nm is held at a height $h_{tip}= 10$~nm above the graphene.  In the measurements presented here, the tip is grounded, so that the charge on the tip is set by the contact potential between the tip and the graphene.  In addition, image charges are created in the tip from impurities on the surface of the graphene sample.  

The spatial profile of the density perturbation created in the graphene by the SPM tip was computed using classical electrostatic finite-element simulations (Maxwell, Ansoft LLC).  The graphene is modeled as a planar conductor, with the observed offset $V_{Dirac}$ of the Dirac point modeled by a homogeneous layer of charge above the graphene.  The tip is realistically shaped and located above the sample at a height $h_{tip}=10$~nm.  The back gate is modeled as an infinite conducting plane, separated from the sample by 280~nm of SiO$_2$. The average carrier density in the graphene is found to be $n = \alpha (V_g - V_{Dirac})$, with $\alpha = 8.3\times10^{10}~\mathrm{cm}^{-2}~\mathrm{V}^{-1}$.  The spatial profile of the image charge created by the SPM tip in the graphene layer has a maximum $\sim 5\times10^{11}~\mathrm{cm}^{-2}$ and a Lorentzian-like shape with half-width at half maximum (HWHM) $r_{scat}\approx 25$~nm. The size and magnitude of the tip perturbation can be compared to the naturally occurring variations in charge density (charge puddles) in graphene, which are found experimentally~\cite{Martin:2008,Deshpande:2009,Zhang:2009} and theoretically~\cite{rossi:2008} to have charge densities $\sim 4\times10^{11}~\mathrm{cm}^{-2}$ with a characteristic diameter $\sim20$~nm, for graphene flakes on a SiO$_2$ substrate.  The perturbation to the charge density created by the SPM tip has approximately the same amplitude, and about double the spatial size of these pre-existing inhomogeneities.   

The conductance fluctuations visible in figure~\ref{fig:setup}b can be identified as UCF.  They are reproducible, and have a root-mean-squared (rms) magnitude $\delta G =  0.64~e^2/h$.  Figure~\ref{fig:setup}c shows the conductance fluctuation $\Delta G$ \textit{vs.} $V_g$ when the tip is fixed far from the sample with tip height $h_{tip}>100~\mu$m.  The three traces from consecutive $V_g$ sweeps show good reproducibility.  Bringing the charged tip near the graphene ($h_{tip}=10$~nm) creates an image charge in the electron gas that significantly alters the conductance fluctuations.  Three $\Delta G$ \textit{vs.} $V_g$ traces in figure~\ref{fig:setup}d for different tip positions spaced $100$~nm apart, demonstrate that UCF is sensitive to the spatial configuration of scatterers -- a change in the position of a single scatterer is enough to decorrelate the conductance fluctuations, as predicted by theory~\cite{Feng:1986, Altshuler:1985b}.

\section{Results}

\subsection{Experimental UCF images}
 \begin{figure}
 \centering
\includegraphics[width=.9\textwidth]{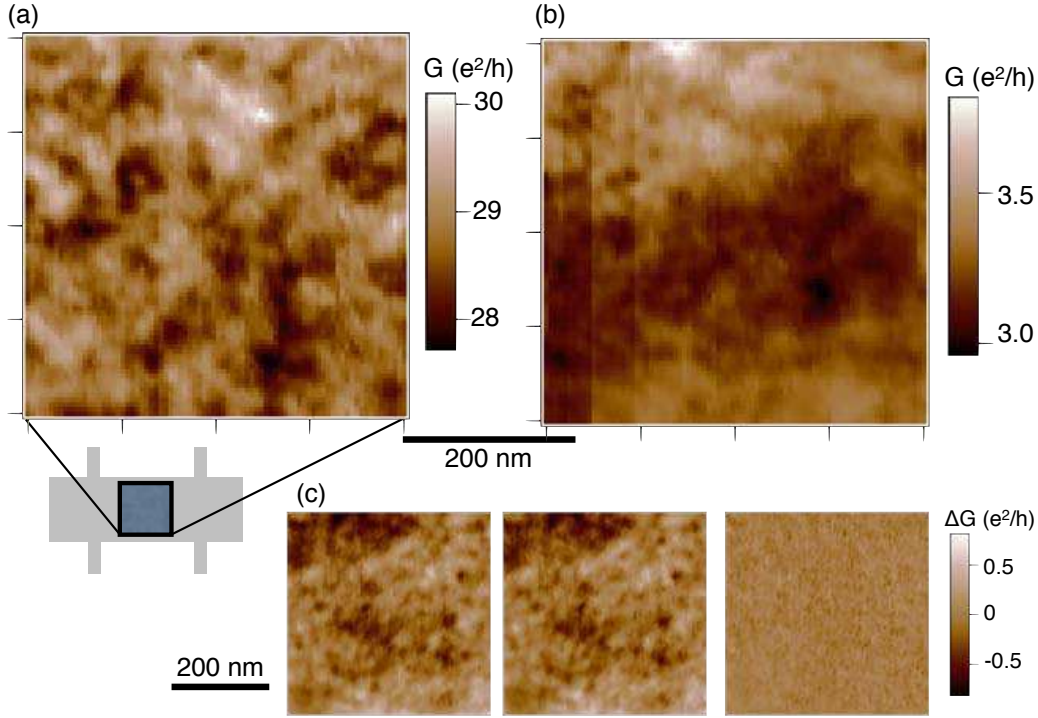}
\caption{(a) Conductance image $G(\mathbf{r})$ \textit{vs.} tip position $\mathbf{r}$ at $T=4$~K, for tip height $h_{tip}=10$~nm.  The density $n = -2.7\times10^{12}$~cm$^{-2}$ ($V_g=-10$~V) is far from the Dirac point.  The $400\times400$~nm$^2$ scan area is located in the center of the sample, as indicated in the schematic diagram. (b) Same as (a) except for a density $n = -1.2\times 10^{11}$~cm$^{-2}$ ($V_g = 20$~V) near the Dirac point. (c) Repeatability is demonstrated by two $400\times400$~nm$^2$ conductance images taken in succession, a few minutes apart, with the rightmost panel showing the difference between the first two images.\label{fig:images}}
 \end{figure}

Using our SPM, we can study UCF by controllably raster scanning the tip position over an area of the sample; previous studies of the effect of single scatterers were based on charge hopping at random positions~\cite{Gusev:1989,Ralls:1993}.  Figures~\ref{fig:images}a and \ref{fig:images}b show conductance images of $G$ \textit{vs.} tip position $\mathbf{r}$ at densities $n = -2.7\times10^{12}$~cm$^{-2}$ and $n = -1.2\times 10^{11}$~cm$^{-2}$ respectively, in a $400\times400$~nm$^2$ area located at the center of the sample.  At high density (figure~\ref{fig:images}a), conductance fluctuations are observed with rms magnitude $\delta G = 0.35 e^2/h$ and characteristic lateral size $\sim 10$s of nm in agreement with UCF theory, as shown below.  At low density, near the Dirac point (figure~\ref{fig:images}b), the conductance fluctuations have a smaller magnitude ${\delta G \approx0.1~e^2/h}$ and a larger lateral size $\sim 100$~nm. Previous transport measurements show the magnitude of UCF in single and multilayer graphene~\cite{Graf:2007,Staley:2008,Horsell:2009} decreases monotonically towards the Dirac point.  The conductance images shown in figure 2 are reproducible, as expected for UCF.  Two images taken $\sim 3$~min. apart (figure~\ref{fig:images}c) are nearly identical; the difference shows only small changes, likely caused by the motion of charged defects in the substrate.  The UCF images shown in figure~\ref{fig:images} are loosely analogous to speckle patterns produced by the coherent scattering of light in a diffusive medium.

\begin{figure}[tbp]
\centering
\includegraphics[width=\textwidth]{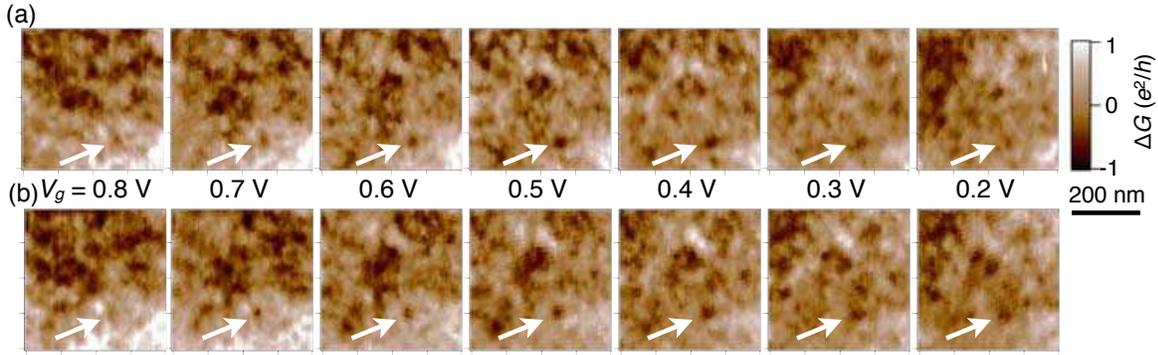}
\caption{(a) Conductance images $G(\mathbf{r})$ \textit{vs.} SPM tip position $\mathbf{r}$ in a $400\times400$~nm$^2$ region of a graphene sample at different back gate voltages $V_g$.  The color scale spans a range of $\pm~1e^2/h$. An arrow points to the same location in each image, highlighting their continuous evolution. (b) A repetition of the series of images in (a), performed 1.5 hrs later, demonstrating the repeatability.  The arrows point to the same location and feature as in part (a).\label{fig:series}}
\end{figure}
Another test for UCF is obtained by measuring the correlation between conductance images recorded at different Fermi energies $E_F$ by changing the density $n$ with the back gate voltage $V_g$. The correlation energy $E_c$ is the range of $E_F$ over which UCF remain correlated. We can determine $E_c$ from our UCF conductance images by finding the change in $V_g$ needed to reduce the correlation between two images by one half; this is the correlation voltage $V_c$.

Figure~\ref{fig:series}a shows a series of conductance images recorded at backgate voltages decreasing from $V_g = 0.8$ to 0.2~V in 0.1~V steps.  By eye, one can see that the images evolve smoothly from one to the next, becoming less correlated as the change in $V_g$ is increased.  For example, the arrows point to the location of a dark spot at $V_g = 0.5$~V in the middle of the series, which has almost completely disappeared at $V_g=0.2$ or $0.8$~V.  The reproducibility of these UCF images over time intervals $\sim 1$~hr is demonstrated by figure~\ref{fig:series}b which shows a repetition of the $V_g$ series, performed 1.5 hrs later.  The arrows point to the same location as in figure~\ref{fig:series}a, showing that the same feature remains.

The correlation $C_{AB}$ between two conductance images $G_A(\mathbf{r})$ and $G_B(\mathbf{r})$ \textit{vs.} SPM tip position $\mathbf{r}$, is $C_{AB}=\int (G_A(\mathbf{r})-\langle G_A \rangle)(G_B(\mathbf{r})-\langle G_B \rangle)d\mathbf{r}$, where angle brackets denote the average over $\mathbf{r}$.  The normalized correlation $\widetilde{C}_{AB}$, such that the autocorrelation of an image is equal to unity is
\begin{equation}
 \widetilde{C}_{AB} = \frac{C_{AB}}{(C_{AA}C_{BB})^{1/2}}.
\end{equation}

From a series of conductance images at different back gate voltages $V_g$, we obtain the normalized correlation $\widetilde C_{(V_g)(V_g+\Delta V)}$ between two images, $G_{V_g}(\mathbf{r})$ and $G_{V_g+\Delta V}(\mathbf{r})$, separated by a fixed change $\Delta V$ in $V_g$.  The average correlation $\langle \widetilde C_{(V_g)(V_g+\Delta V)}\rangle_{V_g}$ \textit{vs.} $\Delta V$ is then obtained by averaging over different values of $V_g$ for a fixed $\Delta V$. 

 \begin{figure}[hbp]
\centering
\includegraphics[width=0.4\textwidth]{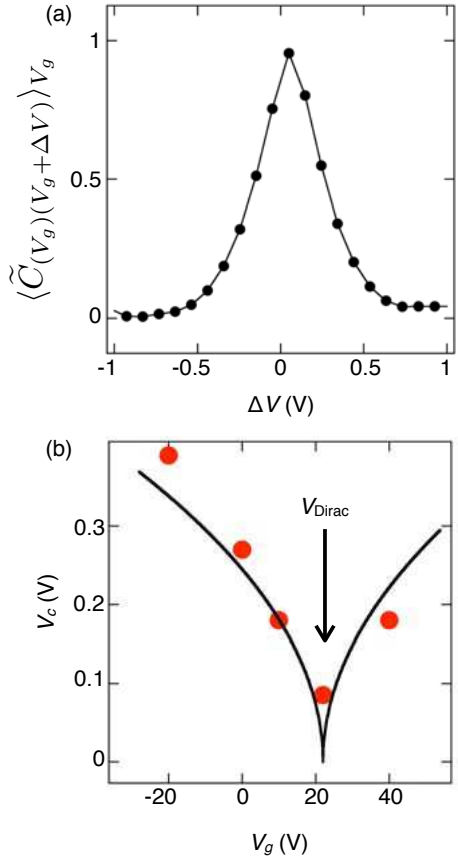}
\caption{(a) Normalized correlation $\langle \widetilde C_{(V_g)(V_g+\Delta V)}\rangle_{V_g}$ \textit{vs.} $\Delta V$ between two UCF conductance images recorded at different back gate voltages $V_g$ and $V_g + \Delta V$, averaged over $V_g$. The correlation voltage $V_c$ is the half-width-at-half-maximum (HWHM) of this curve. (b) Correlation voltage $V_c$ \textit{vs.} $V_g$.  Points: $V_c$ computed from experimental conductance images.  Line: theoretical curve (see text) following Ref.~\cite{Kechedzhi:2009}.\label{fig:vc}}
\end{figure}

Figure~\ref{fig:vc}a shows the normalized correlation $\langle \widetilde C_{(V_g)(V_g+\Delta V)}\rangle_{V_g}$, averaged over $V_g$, \textit{vs.} $\Delta V$ between two UCF images recorded at different back gate voltages $V_g$ and $V_g + \Delta V$ from $V_g = -1.0$ to 1.0~V in steps $\Delta V$~=~0.1~V; the UCF images are similar to those in figure~\ref{fig:series}a.  As shown, the correlation falls off as $\Delta V$ moves away from zero.  The correlation voltage $V_c$ is defined as the halfwidth-at-half-maximum (HWHM) of this curve. To find $V_c$ at a particular back gate voltage $V_g = V_g^0$, a series of conductance images is recorded over the range $V_g=V_g^0-1.0$~V to $V_g=V_g^0+1.0$~V in steps of 0.1~V.  We then calculate $\langle \widetilde C_{(V_g)(V_g+\Delta V)}\rangle_{V_g}$ \textit{vs.} $\Delta V$ for this series of images, and the correlation voltage $V_c$ is obtained from the HWHM of this curve.  For the data in figure~\ref{fig:vc}a we find $V_c=0.27$~V at $V_g=0$.  Note that the range of $V_g$ (2V) covered by this procedure is small compared to the full range of $V_g$  (10s of V) considered in the experiments.

A plot of the measured correlation voltage $V_c$ \textit{vs.} back gate voltage $V_g$ is shown by the red dots in figure~\ref{fig:vc}b.  Theoretically, the correlation voltage ${V_c = (2 E_c /\hbar v_0)(|V_g-V_{Dirac}|/\pi\alpha)^{1/2}}$ where $E_c$ is the correlation energy, and $v_0=1.1\times10^6$~m/s is the Fermi velocity.  The correlation energy is $E_c \approx 2.8~k_B T$ in the thermally broadened regime~\cite{Kechedzhi:2009} that is appropriate here. The theoretical curve for $V_c$ at $T=4$~K, is shown by the solid curve in figure~\ref{fig:vc}b; it is in good agreement with the experimental results providing strong evidence that our conductance images display UCF.
 
To ensure that the measured loss of correlation with increasing change $\Delta V$ in back gate voltage $V_g$ is a repeatable effect and not caused by random drift over the course of the measurement, we calculate the correlation $\widetilde C_{(t)(t^\prime)}$ between a conductance image $G_{V_g,t}(\mathbf{r})$ obtained at time $t$ and the same scan $G_{V_g,t^\prime}(\mathbf{r})$ repeated at $t^\prime = t+1.5$~hrs, as shown in figures~\ref{fig:series}a and \ref{fig:series}b.  The correlation for different times $\langle \widetilde C_{(t)(t^\prime)}\rangle_{V_g} \approx 0.5$, averaged over $V_g$, is much higher than the correlation $\langle \widetilde C_{(V_g)(V_g+\Delta V)}\rangle_{V_g}$ for different back gate voltages separated by $\Delta V = \pm 1$~V, as shown in figure~\ref{fig:vc}a, demonstrating that the complete loss of correlation with gate voltage cannot be caused by drift.

\subsection{Simulated UCF images}
 \begin{figure}
 \centering
\includegraphics[width=0.9\textwidth]{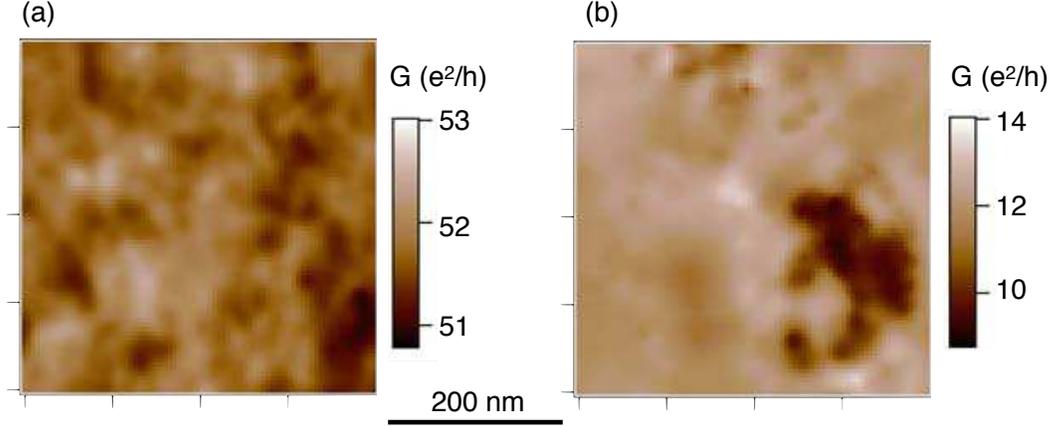}
\caption{(a) Simulated UCF conductance image $G(\mathbf{r})$ \textit{vs.} tip position $\mathbf{r}$ in graphene for a density $n=-8\times 10^{11}$~cm$^{-2}$ away from the Dirac point, and (b) simulated UCF conductance image for a density $n=-4\times 10^{10}$~cm$^{-2}$ near the Dirac point. These simulations show a magnitude $\delta G$ and spatial size of UCF comparable to the data in figure~\ref{fig:images}.  The difference in conductance $G$ between the simulation and experiment is caused by the difference in sample aspect ratio and mobility.\label{fig:sim}}
 \end{figure}
 
We have performed quantum simulations of coherent transport in graphene including the potential from a movable tip~\cite{Topinka:2000}, shown in figure~\ref{fig:sim}, to clarify the origin of features seen in the  measured conductance images.  The simulation results display UCF that change with the tip position, producing conductance images in good agreement with the experiment.  The simulated UCF images show the same trends with $E_F$ as in the experiment.  By varying parameters in the simulation, such as the size of the tip perturbation, we can further explore the effect of a movable scatterer in graphene, and compare to the predictions of analytic theories.
 
Theoretical simulations of UCF conductance images for graphene were obtained by using a finite-difference method to calculate the conductance through a disordered graphene sample~\cite{Tworzydlo:2008}.  The potential in the graphene layer is given by the combination of a local potential from the tip and a number of randomly placed electrostatic scatterers.  Details of the simulations are given in the Appendix.  

By calculating the conductance $G(\mathbf{r})$ \textit{vs.} the tip position $\mathbf{r}$, we produce the same type of conductance images as obtained in the experiments.  Theoretical conductance images, such as figure~\ref{fig:sim}, are obtained by rastering the tip position $\mathbf{r}$ in a plane above the sample. The conductance $G$ at each tip position $\mathbf{r}$ is simulated using the combined potential $U$ from the tip, the intrinsic scatterers and the back gate.  Each image consists of $80\times80$ evaluations of $G$, with tip positions spaced $5$~nm apart, centered within the sample area.  

The simulated images shown in figure~\ref{fig:sim} are in good agreement with the experimental results in figures~\ref{fig:images}a and \ref{fig:images}b.  Figure~\ref{fig:sim}a shows a simulated conductance image at a density $n=-8\times 10^{11}$~cm$^{-2}$ far from the Dirac point.  The image displays spatial conductance fluctuations with amplitude $\delta G\sim e^2/h$, and lateral size $\sim 10$s of nm, similar to the fluctuations observed in the experiment in figure~\ref{fig:images}a.   Close to the Dirac point (figure~\ref{fig:sim}b, $n=-4\times 10^{10}$~cm$^{-2}$), the simulated images show UCF with larger lateral size, in agreement with figure~\ref{fig:images}b.  The simulated UCF images have characteristics similar to the measured images: rms amplitude $\delta G \sim e^2/h$, and spatial size of UCF that increases near the Dirac point.  The agreement between the simulations and the experiment verify that the measured images show UCF caused by the motion of a single scatterer. 

\subsection{Spatially-resolved UCF}

\begin{figure}
 \centering
\includegraphics[width=0.9\textwidth]{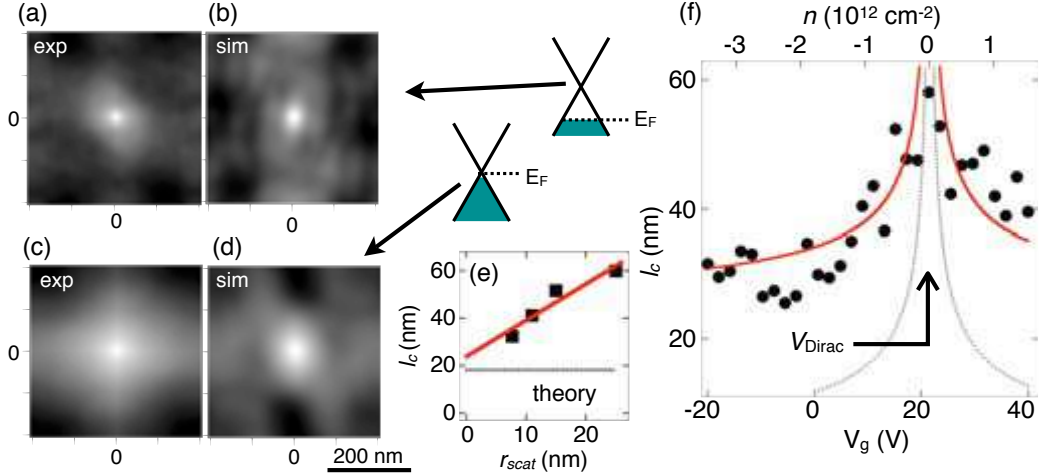}
\caption{(a) Experimental (exp) and (b) simulated (sim) autocorrelation $C(\mathbf{r_0})$ of UCF conductance images away from the Dirac point [(a) $n=-7.1\times 10^{11}$~cm$^{-2}$ and (b) ${n=-8\times 10^{11}}$~cm$^{-2}$].  White = high and black = low.  (c) Experimental (exp) and (d) simulated (sim)  $C(\mathbf{r_0})$ near the Dirac point [(c) $n=-8\times 10^{10}$~cm$^{-2}$ and (d) $n=-4\times 10^{10}$~cm$^{-2}$].  Diagrams of graphene band structure schematically indicate the Fermi energy $E_F$ in (a) to (d).  (e) Simulated correlation length $l_c$ at a fixed $n=-8\times10^{11}$~cm$^{-2}$ \textit{vs.} the radius $r_{scat}$ of the movable scatterer created by the tip, with linear fit (red).  Points are the average of 5 disorder configurations.  Dotted line shows the analytical prediction $l_c=0.46\lambda_F$. (f) Measured correlation length $l_c$ \textit{vs.} $V_g$ (points) obtained from autocorrelations  $C(\mathbf{r_0})$, shown with the analytical prediction $l_c=0.46\lambda_F$ (dotted line), and an empirical fit (red line) to $l_c = 0.46\lambda_F + r_0$ with $r_0=22$~nm; the value of $r_0$ obtained from the fit is close to the SPM tip radius, as expected.  Data points represent the average of four images at slightly different $V_g$ to reduce noise. \label{fig:corrwidth}}
 \end{figure}

Our measurements provide a unique ability to probe theoretical predictions~\cite{Feng:1986, Altshuler:1985b} for the effect of a single movable scatterer on UCF.  To quantify the spatial size of the features in a conductance image $G(\mathbf{r})$, we calculate the spatial autocorrelation $C(\mathbf{r}_0)=\int G(\mathbf{r})G(\mathbf{r} - \mathbf{r}_0)d\mathbf{r}$.  Figures~\ref{fig:corrwidth}a and \ref{fig:corrwidth}b show $C(\mathbf{r}_0)$ away from the Dirac point for the experimental and simulated results, respectively.  The width of the central peak in these plots corresponds to the spatial size of the fluctuations in the original conductance image.  Figures~\ref{fig:corrwidth}c and \ref{fig:corrwidth}d show  $C(\mathbf{r}_0)$ for $E_F$ close to the Dirac point, with significantly broader peaks.

We find that the correlation length $l_c$ for a UCF image is determined by the Fermi wavelength.  The correlation length can be extracted from $C(\mathbf{r}_0)$ by averaging over the angular dependence and defining $l_c$ to be the HWHM of the resulting curve.  Figure~\ref{fig:corrwidth}f shows $l_c$ \textit{vs.} $V_g$ from a series of experimental conductance images spanning the Dirac point.  Theory~\cite{Feng:1986} predicts $l_c \approx 0.46\lambda_F$ (dotted line in figure~\ref{fig:corrwidth}f), where the Fermi wavelength $\lambda_F = 2(\pi/|n|)^{0.5}$.  Both theory and the data show a peak in $l_c$ at the Dirac point.  The analytical calculation, however, does not take into account the spatial extent of the tip or the long-range scatterers (10s of nm) which are not negligible.  

The effect of the spatial size $r_{scat}$ of the movable scatterer created by the tip is investigated in simulated images by finding $l_c$ \textit{vs.} $r_{scat}$, shown in figure~\ref{fig:corrwidth}e, where $r_{scat}$ is the HWHM of the image charge density.  The scatterer effectively smears out the fluctuations on a length scale $(1.4\pm0.3)\times r_{scat}$, shown by the red line in figure~\ref{fig:corrwidth}e.  This smearing can be modeled by adding an offset $r_0$ to the correlation length, $l_c = 0.46\lambda_F + r_0$; a best fit shown by the red line in figure~\ref{fig:corrwidth}f is $r_0 = 22 \pm 1$~nm, which corresponds closely with the tip radius $r_{tip}\approx 20$~nm and with the size $r_{scat}\approx 25$~nm of the image charge created by the tip.

\section{Conclusions}

Our SPM imaging technique probes how coherent transport through a mesoscopic graphene sample is affected by the motion of a single scatterer.  By scanning a charged SPM tip over a graphene device, we obtain conductance images that display fluctuations with amplitude $\delta G \sim e^2/h$ and spatial size $\sim 10$s of nm comparable to the Fermi wavelength.  The UCF conductance images repeat over times up to 1.5 hrs, as predicted for UCF.  The correlation between two images is destroyed by changing the Fermi energy and density, by changing the back gate voltage $V_g$.  The correlation voltage $V_c$ obtained from our measurements agrees well with the theoretical prediction for the correlation energy for UCF.  It is striking that a change $\Delta V_g < 1$~V in back gate voltage is sufficient to completely change the conductance images; this is expected for UCF created by the interference of electron waves traveling along different paths, and demonstrates that the images are not simply reflecting the underlying charge density puddles.  We see that the interference that gives rise to UCF is highly sensitive to the position of even a single scatterer, yielding the full fluctuation $\delta G\sim e^2/h$ when the tip is displaced by only several 10s of nm.  

To verify that the observed conductance images represent UCF caused by the motion of a single scatterer, we simulated the effect of the tip-created scatterer on the conductance of a graphene sample.  The simulated conductance images reproduce the features seen in the experimental images: fluctuations $\delta G\sim e^2/h$ with lateral size $\sim 10$s of nm, which depend sensitively on the Fermi energy and the arrangement of the scatterers.  The simulations also confirm the observed increase in the spatial size of the fluctuations near the Dirac point.

Because universal conductance fluctuations result from quantum interference, one would expect their spatial length scale to depend on the electron wavelength.  Indeed, from the experimental conductance images and numerical simulations, we find that the spatial size of the fluctuations is comparable to the Fermi wavelength $\lambda_F$.  We obtain good agreement with theoretical predictions, taking into account the realistic spatial size of the tip-created scatterer. 

Our measurements demonstrate the utility of a low-temperature scanning probe microscope for studying the coherent flow of electrons through graphene.  The conductance images shown above provide a spatial view of how the interference of electron waves leads to UCF.  This imaging technique will also be useful for the investigation of magnetoconductance fluctuations and weak localization in graphene.  By using a probe of size comparable to the electron wavelength, we gain new insight into the quantum behavior of electrons as they flow through a graphene device.

\ack
We thank K. Brown, H. Trodahl, and E. Boyd for helpful discussions, and acknowledge support from the Department of Energy under grant DE-FG02-07ER46422.  The computations in this paper were run on the Odyssey cluster supported by the Harvard FAS Research Computing Group.

\appendix
\setcounter{section}{1}
\section*{Appendix: Simulation methods}
The numerical calculations we have performed to model our results follow the method described in Ref. \cite{Tworzydlo:2008}.  The sample is discretized into a square lattice and the Dirac equation is solved using a finite difference method on this grid.  An ideal lead is connected to both sides of the sample with propagating electron modes incident on the sample edges.  The conductance is obtained by calculating the transfer matrix for these modes as they travel across the sample.  Further details of these simulations will be given in a separate publication.

In our simulations, the sample grid consists of $102\times153$ points, spaced 5~nm apart.  This places a lower limit on the Fermi wavelength of $\lambda_{min}= 10$~nm, corresponding to a maximum charge density of $n_{max}=4\pi/\lambda_{min}^2=4\pi\times 10^{12}$~cm$^{-2}$.  The direction of current flow is across the narrow dimension of the grid (510~nm), with a width of 765~nm in the transverse direction.  Periodic boundary conditions are applied at the transverse edges, and we focus only on the $400\times400$~nm$^2$ square in the middle to avoid effects of the boundary conditions.  Note that the aspect ratio of the simulated sample $L/W = 2/3$ is less than the aspect ratio $L/W = 2.4$ for the experimental sample.  For the same conductivity, the conductance $G$ for the simulation will be a factor $\simeq 3.6$ larger than $G$ in the experiments. 

Disorder in the graphene is modeled in the simulations as a sum of screened electrostatic potentials created by point charges located above or below the graphene layer.  According to the method of images, a point charge $q$ located a height $a$ above a conducting sheet induces a charge density in the sheet
\begin{equation}
\sigma(\rho)=\frac{-qa}{2\pi(\rho^2+a^2)^{3/2}}
\label{eq:lor}
\end{equation}
where $\rho$ is the radial coordinate away from the position of the point charge.  We then build up the total disorder charge density $\sigma_d(\mathbf{x})$ \textit{vs.} position $\mathbf{x}$ as a sum of such functions, centered at randomly chosen lattice sites, with a fraction $n_i = 0.2$ of lattice sites occupied.  The charge $q=2.5~e$ is chosen to yield a rms charge density $\sigma_d \sim4\times10^{11}~e$/cm$^{2}$ that is in agreement with the observed charge puddles in scanning tunneling~\cite{Zhang:2009,Deshpande:2009} and scanning charge sensor measurements~\cite{Martin:2008}.  The sign of each impurity charge is randomly chosen to be positive or negative, with equal numbers of positive and negative charges.  (The offset of the Dirac point from $V_g=0$ in the experiment is not explicitly included in the simulation.) The distance of the charged impurities from the sample is set to $a=10$~nm to match the lateral size scale of the image charge puddles with puddles observed in scanning tunneling and scanning charge sensor experiments~\cite{Martin:2008,Zhang:2009,Deshpande:2009}.  These numbers combine to yield an effective density $n_{imp} = 2\times 10^{12}$~cm$^{-2}$ of impurities with charge $\pm e$.  The resulting simulated conductance increases linearly with $n$ away from the Dirac point, and is rounded off at $n=0$.  The mobility of the simulated sample is $\mu \simeq 15000$~cm$^2$/V~s, which is a factor of $\simeq 2$ greater than the experimental mobility.  The minimum conductivity is $5.3~e^2/h$ in the simulations, in agreement with the measured value $5.7~e^2/h$. 

We simulate a conductance image by adding to the charge density an additional perturbation $\sigma_{tip}$ created by the tip, centered at position $\mathbf{r}$.  We model the tip as a point charge above the sample, so $\sigma_{tip}$ has the functional form given in equation~\ref{eq:lor}.  We adjust the tip height $a$ to control the width of the tip perturbation $r_{scat}$.  For the images in figure~\ref{fig:sim}, we chose $a=10$~nm, the same distance from the sample as the charged impurities. To test the effect of the radius of the scatterer created by the tip, we vary $a=10$~nm to $a=32.5$~nm in figure~\ref{fig:corrwidth}e.  We set the tip charge $q$ to yield a peak image charge density $\sigma_{max}\sim5\times10^{11}~e/$cm$^2$, as determined from electrostatic simulations, described above.  

Finally, an overall offset $\sigma_0$ to the charge density is added to yield the desired Fermi energy, controlled by $V_g$ in the experiment.  The total charge density in the graphene is then given by $\sigma=\sigma_0+\sigma_d+\sigma_{tip}$.  Using the relationship between the Fermi energy and charge density in graphene, we can now find the potential \textit{vs.} position $\mathbf{x}$ in the graphene layer:
\begin{equation}
U(\mathbf{x})=\hbar v_0\times \mathrm{sgn}(\sigma(\mathbf{x}))\sqrt{\pi|\sigma(\mathbf{x})|}
\label{eq:potential}
\end{equation}
where $v_0$ is the Fermi velocity, and ``sgn'' is the sign function.  This potential is then plugged into the simulation to model the disordered potential through which the electrons flow.  Note that the square root in equation~\ref{eq:potential} means that the different contributions to the potential do not add arithmetically to the total potential.  That is, the contribution to the potential from disorder becomes smaller as the overall charge density increases.  

The empirical model for the disorder described above is based on calculations for charge puddles caused by screened impurities in graphene in Refs.~\cite{galitski:2007,rossi:2008}, and measurements of the size and magnitude of charge puddles in Refs.~\cite{Martin:2008,Zhang:2009,Deshpande:2009}.  We neglect short-range scattering with lattice defects, scattering from ripples or trigonal warping, and the quantum corrections to the screening expected at low density~\cite{galitski:2007}.  We find that this simple model of long-range, ideally-screened electrostatic scatterers is sufficient to reproduce and understand the experimentally observed phenomena.

\end{document}